\PassOptionsToPackage{dvipsnames}{xcolor}
\documentclass[unnumsec,webpdf,contemporary,large]{oup-authoring-template}%
\usepackage{booktabs}

\usepackage{algorithm}
\usepackage{algpseudocode}
\usepackage[mathlines, switch]{lineno}   

\usepackage{tikz}
\usepackage{pgfplots}
\pgfplotsset{compat=1.18}
\usetikzlibrary{decorations.pathmorphing,decorations.pathreplacing,arrows.meta,calc}
\usepackage{etoolbox}
\AtBeginEnvironment{table}{\nolinenumbers}
\AtBeginEnvironment{figure}{\nolinenumbers}
\usepackage{bbding}
\usepackage{multirow}
\usepackage{makecell}
\usepackage{adjustbox}
\usepackage{comment}
\definecolor{DeepRed}{RGB}{180,30,30}

\graphicspath{{Fig/}}
\newcommand{\societylogo}{}

\theoremstyle{thmstyleone}%
\theoremstyle{thmstyletwo}%
\newtheorem{example}{Example}%
\theoremstyle{thmstylethree}%

\onecolumn
\begin{document}

\journaltitle{Journal Title Here}
\DOI{DOI added during production}
\copyrightyear{YEAR}
\pubyear{YEAR}
\vol{XX}
\issue{x}
\access{Published: Date added during production}
\appnotes{Paper}

\firstpage{1}


\title[Short Article Title]{A Configurable Heuristic for the MLCS Problem}

\author[1,$\ast$]{Farhana Akter Tumpa\ORCID{0009-0007-0344-9128}}
\author[1]{Rebin Silva Valan Arasu\ORCID{0000-0002-0573-4893}}
\author[1]{Rajiv Gupta\ORCID{0000-0002-9348-3974}}

\address[1]{\orgdiv{Department of Computer Science}, \orgname{University of California, Riverside}, \orgaddress{\street{900 University Ave, Riverside}, \postcode{92521}, \state{CA}, \country{USA}}}




\corresp[$\ast$]{Corresponding author. \href{email:email-id.com}{ftump001@ucr.edu}}

\received{Date}{0}{Year}
\revised{Date}{0}{Year}
\accepted{Date}{0}{Year}

\abstract{\mdseries
\textbf{Motivation:} 
The Multiple Longest Common Subsequence (MLCS) problem for an arbitrary number of sequences is an NP-hard problem in sequence analysis. Existing exact algorithms based on dynamic programming or MLCS-DAG pruning rapidly exhaust memory as sequence lengths and set sizes grow, while heuristic and hyper-heuristic approaches compromise solution quality and require heavy parameter tuning, respectively. \\
\textbf{Results:} This paper presents the \textbf{ARP} heuristic that, for a given primary sequence, performs three key actions: (i) \textbf{A}dding-$\Delta$s, which incrementally builds a solution by adding subsequences, called $\Delta$s, of the primary sequence to an initial solution;
(ii) \textbf{R}eplacing-Subsequences, which enhances the diversity of the solution set via targeted replacement of subsequences common to all solutions; and  (iii) \textbf{P}rioritizing groups of characters from the primary sequence that are likely to appear together in longer common subsequences. \textbf{ARP} allows the user to \textbf{configure} the degree of Replacements and Prioritization actions for carrying out quality-runtime tradeoff. Empirical evaluations on both synthetic and biological sequence sets demonstrate that \textbf{ARP}'s fastest configuration AOnly finds significantly longer common subsequences than the BNMAS classical heuristic and its aggressive configuration attains solution quality comparable to state-of-the-art hyper-heuristic (UB-HH) while running 1.1$\times$--1.7$\times$ faster.\\
\textbf{Availability and implementation:} The program code is available at \url{https://github.com/ftumpa001/ARP-MLCS} and \url{https://doi.org/10.5281/zenodo.19168760}.\\
\textbf{Contact:} \href{mailto:ftump001@ucr.edu}{ftump001@ucr.edu}
\\
\textbf{Supplementary information:} Supplementary materials are available at \textit{Bioinformatics} online.
}





\maketitle


\section{Introduction}
The Multiple Longest Common Subsequence (MLCS) problem~\cite{maier1978complexity} is widely recognized for its diverse applications, spanning domains such as computational and molecular biology, while also advancing pattern recognition~\cite{lu2007sentence}, string editing, data compression, file plagiarism checking~\cite{miller1985file}, text editing, similarity matching~\cite{needleman1970general}, and information extraction. In the biological context, identifying such subsequences is vital for detecting conserved regions in DNA, RNA, or proteins, which often reflect functional or structural similarities and provide insights into evolutionary processes.  
 Formally, given a set of sequences \(S=\{S_0,S_1,\dots,S_{N-1}\}\) over a finite alphabet \(\Sigma\), the objective of MLCS is to determine the longest subsequence that appears in all sequences. Here, a \emph{subsequence} is obtained by deleting zero or more characters from a sequence without altering the relative order of the remaining characters.

MLCS have been studied for more than three decades. For two sequences, classical dynamic programming computes an exact LCS in \(O(|S_0||S_1|)\) time and space. However, extending dynamic programming to \(N \ge 2\) requires an \(N\)-dimensional table whose size grows exponentially with the number of sequences. To maintain exactness, while reducing this growth, leading exact approaches adopt dominant-point formulations that formulate MLCS as a longest-path problem on a pruned directed acyclic graph (DAG). Exact methods, such as Fast LCS~\cite{chen2006fast}, RLP-MLCS~\cite{li2016real}, Big-MLCS~\cite{wang2021branch}, Quick-DPPAR~\cite{wang2010fast}, BEST-MLCS~\cite{wei2021branch}, and KP-MLCS~\cite{li2023mining}, substantially reduce the search space while preserving optimality.
Among these, KP-MLCS is one of the most advanced but still faces severe runtime and memory demands for long genomic sequences.
However, for long genomic sequences and larger \(N\), even pruned DAGs can expand rapidly as dominant points multiply and the number of distinct MLCS becomes exponential, 
 ultimately exhausting memory and dramatically increasing runtime. Given the above limitation of exact methods, heuristic and meta-heuristic approaches were introduced to efficiently obtain approximate solutions in shorter time for larger sequence sets. 
 While the Long Run~\cite{jiang1995approximation} and Expansion~\cite{bonizzoni2001experimenting} heuristics grow the candidate subsequence step by step, Enhanced Long Run~\cite{huang2004fast} leans on symbol frequencies, BNMAS~\cite{huang2004fast} picks the next symbol that best sets up future extensions, Ant Colony Optimization~\cite{shyu2009finding} follows pheromone-trail search, and beam search~\cite{djukanovic2019beam} keeps a small set of promising partial solutions at each level. More recently, hyper-heuristics such as \textit{UB-HH}~\cite{abdi2023longest} and \textit{LZ-HH}~\cite{nasrollahi2023lempel} have been proposed; they adaptively select base heuristics according to input properties. For example, \textit{UB-HH} employs the set-similarity dichotomizer \(\omega^{2}\varepsilon\) to classify sequence  as correlated or uncorrelated and then selects a heuristic from among \textit{BS-Ex}, \textit{GCoV}, or \textit{kanalytics}~\cite{abdi2022longest}. While these methods improve scalability, they require careful tuning, often dataset-specific, and require significant memory.


In this work, we present \textbf{ARP}, a novel algorithmic formulation inspired by Delta Debugging~\cite{zeller1999yesterday}, which recursively partitions an input and discards non-essential elements to isolate a one-minimal subset. By treating each character in a chosen primary sequence as a "delta" (i.e., a candidate for inclusion), our \textbf{A}ddition algorithm begins with an empty sequence and iteratively adds characters through \emph{subset-based partitions} (S-splits) and \emph{complement-based partitions} (C-splits), modifying granularity when neither can make progress to generate different solutions using different seed solutions. Our \textbf{R}eplacement algorithm further improves upon the generated solutions by removing common subsequences among the solutions and allows the solution sequences to expand in different search directions. Finally, our \textbf{P}rioritization algorithm induces a search ordering which provides guidance to search algorithms in both our Addition and our Replacement algorithms. This helps in finding solutions with longer lengths and in less runtime. 
\textbf{ARP} contributions include:
\begin{itemize}
  \item  Our \textbf{A}ddition \textbf{(AOnly)} algorithm is a simple strategy of character \emph{inclusion maximization} on a chosen primary sequence giving maximal addition set.
  

  \item Our \textbf{A}ddition algorithm uses Subset/Complement-splits with adaptive granularity alleviating the need for tuning. Our \textbf{R}eplacement algorithm's depth and our \textbf{P}rioritization algorithm's beam width parameters can be used to configure the runtime versus quality tradeoff of each algorithm separately.

  \item Our Addition-only algorithm serves as a fast MLCS heuristic that finds significantly longer common subsequences than state of the art BNMAS heuristic with similar runtime.
  Comparison of full blown ARP with UB-Hyper Heuristic shows that ARP achieves similar solution quality but runs 1.1$\times$--1.7$\times$ faster. 
\end{itemize}

\vspace{-0.25in}
\section{The ARP Algorithm}

\noindent\textit{The MLCS Problem:}
Let $S=\{S_0,\ldots,S_{N-1}\}$ be $N$ sequences over an alphabet $\Sigma$ (e.g., $\Sigma=\{A,C,G,T\}$ for DNA).
A sequence $x$ is a subsequence of a sequence $y$ (written $x \preceq y$) if $x$ can be obtained from $y$ by deleting zero or more symbols while preserving the order of symbols.
The Multiple Longest Common Subsequence (MLCS) problem asks for a sequence $x^\star$ of maximum length that is a subsequence of every $S_j$ ($j=1,\ldots,N$).
When multiple maximizing subsequences exist, returning any one is acceptable; their common length is denoted $L^\star$. Therefore:

\[
x^\star \in \operatorname*{arg\,max}_{x}\Big\{\,|x| \;:\; x \preceq S_j \ \text{for all } j=0,\ldots,N-1 \Big\}.
\]

\noindent
As an illustration, consider three DNA sequences: \textit{S$_0$ = AGACTACT, S$_1$ = ATCCTGCT, S$_2$ =AACTGTCT}. The subsequence \textit{ACTCT} occurs in all three after deleting different symbols while preserving order, and, since no longer common subsequence exists, it represents the multiple longest common subsequence with length $L^\star=5$.

\vspace{0.05in}
In this paper, we reformulate the MLCS problem as a process of \textit{character selection} from a chosen sequence. 
Given a set of input sequences $S = \{S_0, S_1, \ldots, S_{N-1}\}$, one sequence $P \in S$ is selected as the \emph{primary}. Then, MLCS can be defined as the longest subsequence of the primary sequence that is also a subsequence for all other sequences. First, in our formulation, we explore search space constructively rather than reductively. Second, we enhance the diversity of the solution set via targeted replacement of subsequences common to all solutions. Third, we order addition of characters such that characters likely to be present in longer common subsequences are added first.

\vspace{0.1in}
\noindent
\textit{The ARP Algorithm: }
The \textbf{ARP} heuristic, for a given a primary sequence, perform three kinds of actions: 
\begin{itemize}\itemsep2pt
\item
\textbf{A}dding-$\Delta$s, which incrementally build a solution by adding subsequences, called $\Delta$s, of the primary sequence to an initial solution. Adding $\Delta$s to alternate initial solutions produces a solution set. The $\Delta$ representation encodes each position in the primary sequence as a potential selection candidate, establishing the foundation for iterative maximization. The granularity of additions is varied, beginning with coarse-grained additions of larger sequence segments to finer-grained which finally ends with individual characters. Two complementary forms of \textit{splits} regulate how the search space is partitioned and traversed;
\item
\textbf{R}eplacing-Subsequences, which enhances the diversity of the solution set via targeted replacement of subsequences common to all solutions. The replacement identifies the subsequence common to all solutions found in the addition process and removes them from each solution to create a diversified solution set; and 
\item 
\textbf{P}rioritizing the indices of character from the primary sequence allows the Addition and Replacement algorithms to proceed faster as groups of correlated character can be added or removed together. Prioritization also leads the ARP algorithm to find better solutions because the indices that are expected to lead to smaller maximal solutions are visited last.

\end{itemize}

Finally, \textbf{ARP} incorporates parameters that allow the user to \textbf{configure} the degree of Replacements and Prioritization actions for carrying out quality-runtime tradeoff.
\subsection{Definitions: $\Delta$ Representation and Splits}

Before we dive in into our algorithm, we first define the preliminaries for our work. To explain the preliminary concepts, we will be using the following example. 

\vspace{-0.1in}
\begin{example}
\label{example:1}

Consider three sequences 
\( S_0 = \texttt{AGACTACT} \),
\( S_1 = \texttt{ATCCTGCT} \),
and 
\( S_2 = \texttt{AACTGTCT} \),
with \( P = S_0 \) as the primary sequence.
\end{example}

\vspace{-0.1in}
\textbf{$\Delta$ Representation: }
Let \( P = p_{0}p_{1}\ldots p_{|P|-1} \) be the \textit{primary sequence} selected from the input set \( S=\{S_{0}, S_{1}, \ldots S_{N-1}\} \) and an index set $\Delta_P$ is any \textit{ordered} subset of indices in $P$, $\Delta_P \subseteq \{0, 1, \ldots, |P|-1\}$. For any index set $\Delta_P = \{\delta_0, \delta_1, \ldots \delta_{|\Delta_P|-1}\} \subseteq \{0, \ldots, |P|-1\}$, the subsequence $p_{\delta_{i_0}}p_{\delta_{i_1}}...p_{\delta_{i_{|\Delta_P|-1}}}$ (where $\forall j, (\delta_{i_j} \in \Delta_P) \land (\delta_{i_j} < \delta_{i_{j+1}})$ i.e, subsequence obtained by indexing the primary string after reordering the index set in ascending order) is obtained by concatenating the characters $p_{\delta_i}$ at positions $\delta_i$ in $P$ after arranging the $\delta_i$ in ascending order. In our example, the index set of $P$ with all indices is $\{0, 1, \ldots 7\}$. For ease of interpretation, henceforth, we will represent the index set in the following format: $A^0G^1A^2C^3T^4A^5C^6T^7$ 
, representing the potential addition of the \( \delta_i^{\text{th}} \) character $p_{\delta_i}$ from \( P \) to current solution. Note that the index set is ordered but not necessarily in ascending order (can be seen in the example below).

This representation forms the basis for our formulation of the MLCS problem, where each index \( \delta_{i} \) also serves as a candidate $\delta_i$ for addition of character $p_{\delta_i}$ to the current solution subsequence represented by the Current Index Set (CIS). We further define the following operations on Index sets. For the following operations, consider the example with index sets $A_P = \{2, 4, 1, 6\} \equiv A^2T^4G^1C^6 $ and $B_P = \{7, 5, 1, 4\} \equiv T^7A^5G^1T^4$.

\begin{enumerate}
    \item \textbf{Union}: Given two Index sets $A_P$ and $B_P$ on the same primary sequence $P$, we define $A_P \cup B_P$ as the ordered union that contains all the indices of $A_P$ and then the indices of  $B_P$. In our example, $A_P \cup B_P = \{2, 4, 1, 6, 7, 5\} \equiv A^2T^4G^1C^6T^7A^5$.
    \item \textbf{Intersection}: Similarly, given two Index sets $A_P$ and $B_P$ on the same primary sequence $P$, we define $A_P \cap B_P$ as the ordered intersection that contains indices of that are present in both $A_P$ and $B_P$ preserving the order from $A_P$. In our example, $A_P \cap B_P = \{4, 1\} \equiv T^4G^1$.
    \item \textbf{Set Difference}: Given two Index sets $A_P$ and $B_P$ on the same primary sequence $P$, we define $A_P \setminus B_P$ as the indices of $A_P$ that are not in $B_P$ preserving the order from $A_P$. In our example, $A_P \setminus B_P = \{2, 6\} \equiv A^2C^6$. 
\end{enumerate}

For two index sets $A_P$ and $B_P$ on the same primary sequence $P$, we also define the addition operator $A_P \oplus B_P$ as any ordered maximal superset of $A_P$ that is a common subsequence of all sequences in $S$ containing only elements from $A_P$ and $B_P$. In other words, $A_P \oplus B_P$ is a maximal set such that $A_P \subseteq A_P \oplus B_P \subseteq A_P \cup B_P$ and is also a valid common subsequence of all sequences in $S$. The ordering of $A_P \oplus B_P$ is that it first contains all elements of $A_P$ and contains elements of $B_P$ in the same order of $B_P$. Our  $A_P \oplus B_P$ implementation is shown later in Algorithm \ref{alg:add-mlcs-generator}.

Our ARP algorithm consists of three functions presented in subsequent sections: \textsc{Replacement} promotes variability among solutions; {\sc Addition} finds a solution by adding; and {\sc Reorder} prioritizes characters in the primary sequence for {\sc Addition} and {\sc Replacement}.

\subsection{The \textsc{Replacement} Algorithm }

\begin{algorithm}[H]
\caption{The {\sc Replacement} Algorithm.}
\label{alg:rec-mlcs-generator}
\begin{algorithmic}[1]
\vspace{4pt}
\Require $P$: a sequence from $S$ on which the solution is searched.
\Ensure $MLCS$: a maximal subsequence of $S$
\Procedure {\sc Replacement }{$soln_P, activeIndices, depth$}
    \State {\sc Solns} $\gets$ $\emptyset$
    \State $seed_P \gets soln_P$
    \While{\Call{Check}{$seed_P$}}
        \State $soln_P \gets seed_P \bigoplus activeIndices$ \label{algo:line:setsoln}
        \State $seed_P \gets seed_P \; \bigcup \; (activeIndices_P \setminus soln_P)$ 
        
        \Comment{Find new solutions including complement of current}
    
        \State $solns \gets  solns \; \bigcup \; \{soln_P\}$
    \EndWhile

    \State $cs_P \gets \bigcap\limits_{soln_P \in solns} soln_P$
    
    \Comment{Find subsequence common to all solutions}

    \vspace{0.5em}
    \If{depth $<$ MAX}
    
    \Comment{MAX depth parameter controls the recursion depth}
        \For{$soln_P \in solns$}
            \State $replaced_P \gets (soln_P \setminus cs_P) \bigoplus (activeIndices_P \setminus soln_P)$
            
            \Comment{Replace Common Subsequence in soln} \label{algo:line:replace}
            
            \State $new\_soln_P \gets \Call{Replacement}{cs_P,  replaced_P, depth+1}$
            
            \Comment{Improve new solution} \label{algo:line:addback}
            
            \If{$|new\_soln_P| > |soln_P|$}
                $soln_P \gets new\_soln_P$ 
                
            \EndIf
        \EndFor
    \EndIf
    \State \Return Best of $solns$
\EndProcedure

\end{algorithmic}
\end{algorithm}

Addition-based MLCS algorithms that iteratively adds characters or groups of characters to its solution guarantees maximality of its solution when it is done. Maximal solutions are subsequences that cannot be made any longer by adding characters, In usual addition based MLCS algorithms, one of the main problems is that most of the solutions converge to the same subsequence due to the presence of certain groups of characters which drive maximal solutions to the same subsequence. To mitigate this, our replacement algorithm focuses on subsequences common to all current solutions and tries to replace them with other characters. This brings in a variance within the current set of solutions resulting a better exploration of the search space.

As can be seen from Algorithm \ref{alg:rec-mlcs-generator}, the replacement algorithm creates a set of solutions using our addition algorithm with different initial solutions (seeds) in Line \ref{algo:line:setsoln}. The first seed is just the empty sequence and subsequent seeds contain the indices  not in previous solutions, ensuring each solution is unique. 
Moreover, our replacement algorithm finds the subsequence common to the entire set of solutions, removes them from all solution  and expands each solution using our addition algorithm with characters that were not previous used as can be seen from Line \ref{algo:line:replace}. If successful, this brings in much more variability among the set of solutions. Finally, the previously removed characters are added back to each solution recursively, if they could still be added, to ensure maximality (Line \ref{algo:line:addback}). This will not restore any modification done by replacement due to the maximality guarantee unless no character was ever replaced. Since at most $|P|$ solutions can be found in Line~\ref{algo:line:setsoln}, the branching factor is $|P|$. Therefore, the time complexity of the algorithm is at most $O(|P|^{MAX} T(|P|))$, where $T(|P|)$ is the time complexity of the addition algorithm.

For example, consider the binary sequences and the primary sequence shown below:

\vspace{-0.1in}
\begin{center}
\textit{Sequences}: "\texttt{100111100100}", "\texttt{100010110010}", and "\texttt{100111100010}" \\ \textit{Primary Sequence}: "\texttt{100111100100}"
\end{center}

\begin{figure}[t]
\centering
\includegraphics[width=\columnwidth]{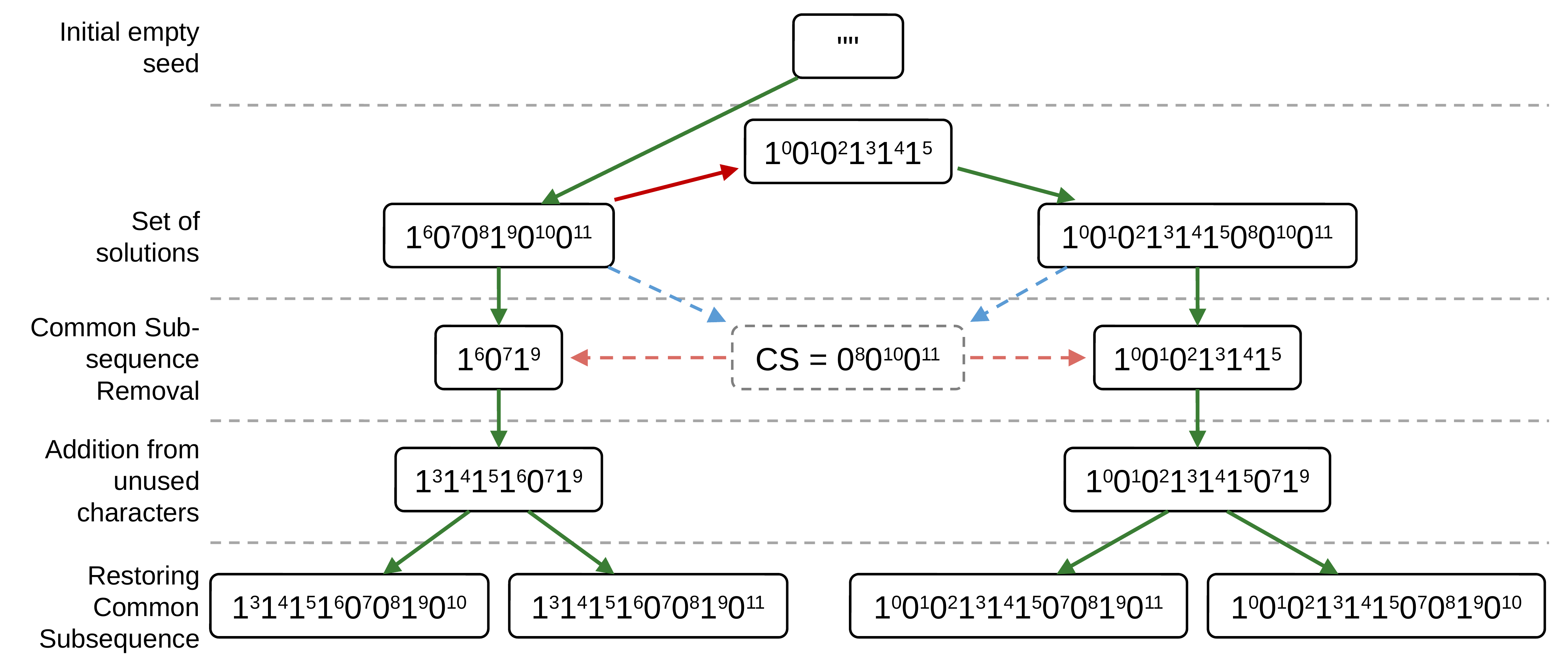}
\caption{Progression of Replacement algorithm for  ($|\Sigma|=2$) sequences $S=\{"100111100100", "100010110010", "100111100010"\}$ \& $P=S_0$.}
\label{fig:acr}

\vspace{-0.15in}
\end{figure}

\noindent
The complete index set of the Primary sequence is $1^00^10^21^31^41^51^60^70^81^90^{10}0^{11}$. As shown in Figure \ref{fig:acr}, the addition algorithm generates two solutions $soln_{P1} = 1^60^70^81^90^{10}0^{11}$ with empty seed and $soln_{P2} = 1^00^10^21^31^41^50^80^{10}0^{11}$ with seed $\{0..11\} \setminus soln_{P1} = 1^00^10^21^31^41^5$. The \textsc{Replacement} algorithm identifies the common subsequence between the two solutions as $cs_P = soln_{P1} \cap soln_{P2} = 0^80^{10}0^{11}$. Then, $soln_{P1} \setminus cs_P$ and $soln_{P2} \setminus cs_P$ are then expanded using their respective unused indices with the \textsc{Addition} algorithm as $1^31^41^51^60^71^90$ and $1^00^10^21^31^41^51^60^71^9$. The \textsc{Replacement} algorithm is then recursively applied on the expanded sequences to add back the common subsequence to create maximal solutions. The two solutions $soln_{P1}$ and $soln_{P2}$ at the start of the algorithm have lengths $6$ and $9$ which are improved to lengths $8$ and $10$, respectively by the \textsc{Replacement} algorithm.
\subsection{The \textsc{Addition} Algorithm}
\label{section:Addition}

\algnewcommand{\LineComment}[1]{\State \(\triangleright\) #1}
\begin{algorithm}[htbp]
\caption{The {\sc Addition} Algorithm and the $\bigoplus$ operator.}
\label{alg:add-mlcs-generator}
\begin{algorithmic}[1]

\Require $P$ is a sequence from $S$ on which the solution is searched
\Ensure Find a maximal subsequence of $P$ common to all sequences in $S$

\Procedure {\sc $\bigoplus$ \;}{$initial\_soln_P, activeIndices_P$}
    \State $n\_splits \gets 2$ 
           \Comment{Initial granularity or the number of partitions}
    \State $CIS_P \gets initial\_soln_P$ 
           \Comment{Current Index Set for current solution}
    \State assert $CIS_P \subseteq activeIndices$ 
    

    \While{True}
        \State $subsets \gets$ \Call{split}{$activeIndices_P - CIS_P, n\_splits$} 
               
               \Comment{Split remaining indices into $n\_splits$ subsequences}

        \LineComment{\textbf{C-split}} \label{dd:line:csplit}
        \For{each $subset_P$ in $subsets$}
            \If{\Call{check}{$activeIndices_P - subset_P$}} 
                
                \State $CIS_P \gets activeIndices_P - subset_P$ 
                       
                       \Comment{Expand current subsequence with new candidate}
                
                \State $n\_splits \gets 2$ \Comment{Reset granularity after successful expansion}

                \State \textbf{continue while-loop}
            \EndIf
        \EndFor
        \vspace{1em}
        \LineComment{\textbf{S-split}} \label{dd:line:ssplit}
        \For{each $subset_P$ in $subsets$}
            \If{\Call{check}{$CIS_P \cup subset_P$}} 
                \State $CIS_P \gets CIS_P \cup subset_P$ 
                \State $n\_splits \gets n\_splits - 1$ 
                \State \textbf{continue while-loop}
            \EndIf
        \EndFor


        \vspace{1em}
        \If{$n\_splits = |activeIndices_P| - |CIS_P|$} \\
            \hspace{0.5in} \Return \Call{subsequence}{$P, CIS_P$} 
        \EndIf
                   \Comment{Return the constructed MLCS subsequence}
        \vspace{1em}
        \LineComment{\textbf{Split factor doubling}} \label{dd:line:factor}
        \State $n\_splits \gets \min(2 \times n\_splits, |activeIndices_P| - |CIS_P|)$ 
               
               \Comment{If no progress, improve granularity}
    \EndWhile
\EndProcedure

\end{algorithmic}
\end{algorithm}
The addition algorithm is the core of ARP, growing a maximal solution from an initial index set. The Replacement algorithm was able to be built on top of this algorithm only because of its guarantee of maximality. ARP could theoretically be implemented on any addition-based MLCS algorithm, but our addition algorithm's superior runtime and sensitivity to initial ordering make it tailor-made for ARP.

The addition usually begins with the primary sequence \(P\) represented as a set of potential addition candidates 
\(X_P = \{\delta_{0}, \delta_{1}, \ldots, \delta_{|X_P|-1}\}\) to the initial solution represented by the Current Index Set (CIS). But generally, the addition algorithm can start with any index set of $P$ as the search space.
Initially, these candidates are divided into large groups. The goal at this stage is to add as many characters as possible to the initial solution while ensuring that the CIS remains common across all input sequences in \(S\). Slowly, as search slows down, the search is made finer to keep adding indices until no more index could be added.

To efficiently evaluate candidates during the search process,
the algorithm recursively partitions the current unused set of characters
\( X_P = \{0, 1, \ldots |P|-1\} \setminus CIS_P =\{\delta_{0}, \delta_{1}, \ldots, \delta_{|X_P|-1}\} \)
into smaller subsets $X_{P0}, X_{P1}, \ldots X_{P(n-1)}$ called \textit{splits}.
Two complementary split operations are defined as follows. For the below splits, we will consider the example in Table~\ref{tab:addition_dd_trace3}, $CIS_P = A^0G^1$ and unused set of indices $X_P = A^2C^3T^4A^5C^6T^7$ with $n=3$. Therefore, $X_{P0} = A^2C^3, X_{P1} = T^4A^5$ and $X_{P2} = C^6T^7$
\begin{itemize}
    \item \textbf{Complement Split (C-Split ):} tests whether including all but one subset of $X_P$  preserves commonality (Line~\ref{dd:line:csplit}). If $CIS_P \cup (X_P \setminus X_{Pi}) = \{0, 1, \ldots, |P|-1\} \setminus X_{Pi}$ fails to preserve commonality for all $0 \le i < n$, then the algorithm proceeds to S-split. In the example, the complement splits would therefore be $A^0G^1T^4A^5C^6T^7$, $A^0G^1A^2C^3C^6T^7$ and $A^0G^1A^2C^3T^4A^5$
    \item \textbf{Subset Split (S-Split):} tests whether adding a specific subset  $CIS \cup X_{Pi}$ preserves the common-subsequence property (Line~\ref{dd:line:ssplit}). In the example, the S-splits would therefore be $A^0G^1A^2C^3$, $A^0G^1T^4A^5$ and $A^0G^1C^6T^7$.
\end{itemize}When both subset and complement additions fail to  preserve commonality, the split factor is increased to $n \leftarrow \min(2n, |X|)$ which doubles the number of partitions (Line~\ref{dd:line:factor}), dividing the unused indices set into smaller regions.
As the process continues, the number of partitions \(n\) gradually grows, the search shifts from broad additions to detailed testing of smaller character groups, keeping the search efficient. Through S-Split, C-Split, and adaptive granularity control, the addition algorithm converges to a maximal solution to which no additional index can be added.

Assuming that all sequences are of equal length, the check function can run in $O(N|P|)$ time. Since each C-split and S-split increases the $CIS_P$ by atleast one, there can only be atmost $O(|P|)$ successful C-split and S-split. Also, there can only be $O(log(|P|))$ non-successful splits between two successful splits due to granularity. Accounting for different granularity, there can only be $O(|P|)$ checks between two non-successful splits. This implies that there will be atmost be $O(|P|^2)$ checks in total. Hence, the time complexity of the addition algorithm is $O(N|P|^3)$.
\begin{table*}[!t]
\caption{Step-by-step trace of {\sc Addition} on sequences $P=S_0 =$ AGACTACT, $S_1 =$ ATCCTGCT, and $S_2 =$ AACTGTCT. $\notin S_1$ indicates that the candidate is not a subsequence of $S_1$.}
\label{tab:addition_dd_trace3}
\centering
\renewcommand{\arraystretch}{1.15}
\setlength{\tabcolsep}{1.5pt}
\begin{adjustbox}{max width=\linewidth}
\begin{tabular}{|c|c|c|c|c|c|c|c|}
\hline
\textbf{Run} & \textbf{Current} & \textbf{Remaining} & \textbf{Granularity} & \textbf{Phase} & \textbf{Current} & \textbf{Current} & \textbf{Check} \\
 & \textbf{Solution} & \textbf{Characters} &  & & \textbf{subset} & \textbf{Candidate} &  \\
\hline \hline
1 & "" & $C^6T^4A^0C^3T^7A^2G^1A^5$ & 2 & C-split & $C^6T^4A^0C^3$ & GAAT & $\notin S_1$ \\ 
  &  &  &  &  & $T^7A^2G^1A^5$ & ACTC & \textbf{Pass} \\ 
\hline
2 & $A^0C^3T^4C^6$ & $T^7A^2G^1A^5$ & 2 & C-split & $T^7A^2$ & AGCTAC & $\notin S_1$ \\ 
  &  &  &  &  & $G^1A^5$ & AACTCT & $\notin S_1$ \\ 
  &  &  &  & S-split & $T^7A^2$ & AACTCT & $\notin S_1$ \\ 
  &  &  &  &  & $G^1A^5$ & AGCTAC & $\notin S_1$ \\
  &  &  & 4 & Split factor doubled &  &  &  \\
  &  &  &  & C-split & $T^7$  & AGACTAC & $\notin S_1$ \\
  &  &  &  &  & $A^2$  & AGCTACT & $\notin S_1$ \\
  &  &  &  &  & $G^1$  & AACTACT & $\notin S_1$ \\
  &  &  &  &  & $A^5$  & AGACTCT & $\notin S_1$ \\
  &  &  &  & S-split & $T^7$  & \textcolor{red}{ACTCT} & \textbf{Pass} \\
\hline
3 & $A^0C^3T^4C^6T^7$ & $A^2G^1A^5$ & $4 \rightarrow 3$ & C-split & $A^2$ & AGCTACT & $\notin S_1$ \\ 
  &  &  &  &  & $G^1$  & AACTACT & $\notin S_1$ \\
  &  &  &  &  & $A^5$  & AGACTCT & $\notin S_1$ \\
  &  &  &  & S-split & $A^2$  & AACTCT & $\notin S_1$ \\
  &  &  &  &  & $G^1$  & AGCTCT & $\notin S_1$ \\
  &  &  &  &  & $A^5$  & ACTACT & $\notin S_1$ \\
\hline
\end{tabular}
\end{adjustbox}
\end{table*}


\sloppy To illustrate the workings of our addition algorithm, we will show how our addition algorithm proceeds on Example \ref{example:1} with initial ordering $C^6T^5A^0C^3T^7A^2G^1A^5$. From Table~\ref{tab:addition_dd_trace3}, we can see that  the search starts with an empty sequence which is always a common subsequence. For the C-splits and S-splits, if the split is a common subsequence, then we keep the split for final solution and splits the remaining characters. In the first run, we see that one of the C-splits $T^7A^2G^1A^5$ representing "ACTC" is a valid subsequence of all the sequences. Hence, this index set is chosen as CIS and the search continues on remaining indices. When no C- or S- splits provide a valid solution, the split factor is doubled to search for more finer additions as seen in run 2. Finally, when no splits provide a valid solution and when the split factor can no longer be increased, the search terminates returning the CIS as can be seen in run 3. In Table~\ref{tab:addition_dd_trace3}, $\notin S_1$ indicates that the candidate is not a subsequence of $S_1$ = ATCCTGCT.

\begin{algorithm}[!t]
\caption{The {\sc Reorder} Algorithm.}
\label{alg:bgd_full}
\begin{algorithmic}[1]
\Require $P$: Primary sequence, $S$: set of input sequences, $W$: beam width, $T$: top paths
\Ensure Approximate longest common subsequence shared by all sequences in $S$

\Function{\sc Reorder }{$allIndices_P$}
    \State Compute suffix–LCS tables $L^{(k)}[p,q]$ for all $S_k \in S \setminus \{P\}$
        
        \Comment{Used for tight upper–bound estimation}
    \State Initialize beam $\mathcal{B}$ with an empty subsequence and cursors 
    \For{each position $i$ in $P$}
        \State Expand $\mathcal{B}$ with two branches: \textbf{ADD}($P[i]$) and \textbf{SKIP}($P[i]$)
        \For{each beam state $s = (\ell, i, c)$}
            \State Compute the beam score:$\;\;$  \\
            $\;\;\;\;\;\;\;\;\;\;\;\;\;\;\;\;$ $f(s) = \ell + UB(i, c), \;\;
            where \;\;
            UB(i, c) = \min_{k} L^{(k)}[\,i,\, c_k{+}1\,]
            $
            
            \Comment{$UB(i,c)$ is the maximum remaining match potential across all $S_i$s}
        \EndFor
        \State Retain only the top–$W$ beam states with the highest $f(s)$ values
    \EndFor
    \State Compute survival frequency $f_i$ for each $i$ from the top–$T$ beam paths
    \State Sort $allIndices_P$ in descending order of $f_i$      
        \Comment{High survival indices are added first}
    \State \Return $allIndices_P$
\EndFunction

\vspace{1em}
\Procedure {\sc ARP }{}
    \State $P \gets$ \Call{PrimarySelect}{S}
    \State $allIndices_P \leftarrow$ {\sc Reorder }$(\{0, 1, \ldots, |P|-1\})$ 
    \State \Return \Call {Replace} {$\emptyset, allIndices, 0$}
\EndProcedure

\end{algorithmic}
\end{algorithm}

\subsection{The Reorder Algorithm for Prioritization}In our \textsc{Addition} and \textsc{Replacement} algorithms, character indices are added or deleted based on their order of appearance or frequency, which keeps the process generic but restricts exploration. Since the Addition algorithm systematically tests subsets, index \textit{ordering} directly influences how the algorithm explores its search space. A poor ordering wastes time on unpromising candidates and converges early to suboptimal solutions. The algorithm should focus more on those positions that are consistently useful in forming long, valid common subsequences before spending effort on weaker or inconsistent ones. To solve this, in Algorithm \ref{alg:bgd_full} each index $\delta_i$ in the primary sequence $P$ is assigned a priority score $s_i$ that measures its contribution to high-quality subsequences. A high $s_i$ indicates that the corresponding character frequently appears in long subsequences shared by all sequences, so it should be added early. A low $s_i$ means that the character seldom contributes meaningfully and should be added late. Thus, a better ordering is simply one in which the indices are arranged according to descending priority scores $s_i$. Beam-guided ordering overcomes this bottleneck by statistically learning the importance of each character, directing the algorithm toward the most promising parts of the sequence.

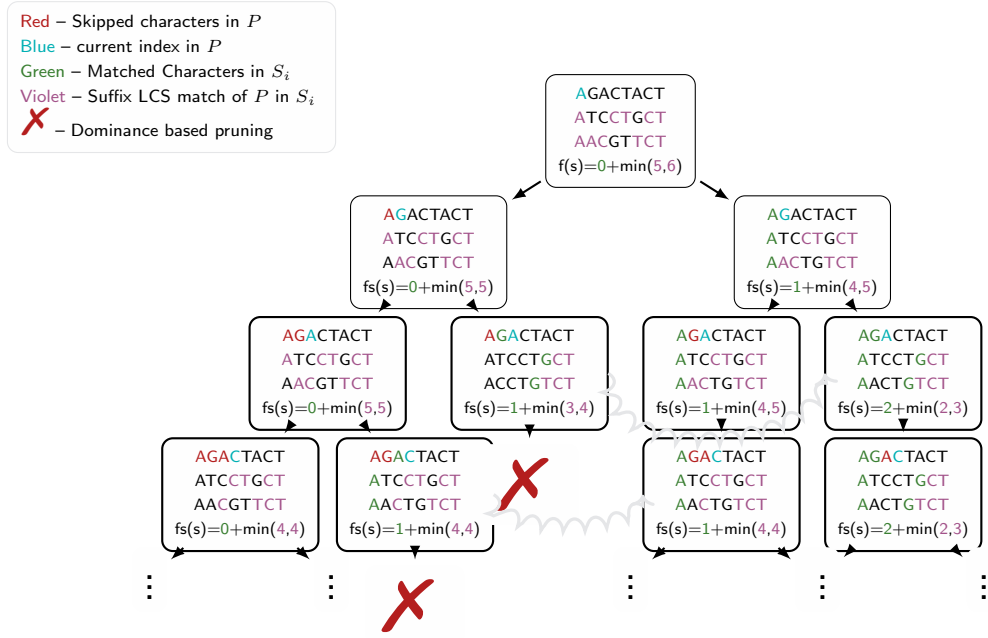
\begin{figure*}[!t]
\centering
\begin{tikzpicture}[scale=0.95, transform shape,
    level distance=6em,
    sibling distance=19em,
    every node/.style={
        draw,
        rounded corners,
        align=center,
        font=\sffamily\footnotesize,
        fill=gray!3,
        inner sep=5pt,
        minimum width=7em
    },
    edge from parent/.style={
        draw,
        thick,
        ->,
        >=latex,
        shorten >=2pt,
        shorten <=2pt
    },
    decoration={coil, amplitude=0.8mm, segment length=2.8mm},
]

\node (seed) [font=\sffamily\small, text width=15em, align=left, draw=gray!70] at (-6.25, 0.75) {
\textcolor{DeepRed}{Red} -- Skipped characters in $P$ \\
\textcolor{TealBlue}{Blue} -- current index in $P$ \\
\textcolor{OliveGreen}{Green} -- Matched Characters in $S_i$ \\
\textcolor{DarkOrchid}{Violet} -- Suffix LCS match of $P$ in $S_i$ \\
{\large \textcolor{DeepRed}{\XSolidBrush}} -- Dominance based pruning
};

\node (root){\textcolor{TealBlue}{A}GACTACT\\\textcolor{DarkOrchid}{A}TC\textcolor{DarkOrchid}{CT}G\textcolor{DarkOrchid}{CT}\\\textcolor{DarkOrchid}{AAC}GT\textcolor{DarkOrchid}{TCT} \\ f(s)=\textcolor{OliveGreen}{0}+min(\textcolor{DarkOrchid}{5},\textcolor{DarkOrchid}{6})}
    child {node (L1) {\textcolor{DeepRed}{A}\textcolor{TealBlue}{G}ACTACT\\\textcolor{DarkOrchid}{A}TC\textcolor{DarkOrchid}{CT}G\textcolor{DarkOrchid}{CT}\\A\textcolor{DarkOrchid}{AC}GT\textcolor{DarkOrchid}{TCT} \\ fs(s)=\textcolor{OliveGreen}{0}+min(\textcolor{DarkOrchid}{5},\textcolor{DarkOrchid}{5})}
        [sibling distance=10em]
        [level distance=6em]
        child {node (L2a) {\textcolor{DeepRed}{AG}\textcolor{TealBlue}{A}CTACT\\\textcolor{DarkOrchid}{A}TC\textcolor{DarkOrchid}{CT}G\textcolor{DarkOrchid}{CT}\\A\textcolor{DarkOrchid}{AC}GT\textcolor{DarkOrchid}{TCT} \\ fs(s)=\textcolor{OliveGreen}{0}+min(\textcolor{DarkOrchid}{5},\textcolor{DarkOrchid}{5})}
            [sibling distance=8.65em]
            child {node (L3a) {\textcolor{DeepRed}{AGA}\textcolor{TealBlue}{C}TACT\\ATC\textcolor{DarkOrchid}{CT}G\textcolor{DarkOrchid}{CT}\\AA\textcolor{DarkOrchid}{C}GT\textcolor{DarkOrchid}{TCT} \\ fs(s)=\textcolor{OliveGreen}{0}+min(\textcolor{DarkOrchid}{4},\textcolor{DarkOrchid}{4})}
                [level distance=4.2em]
            [sibling distance=9em]
                child {node [draw=none,minimum width=1em] {\large \textbf{\vdots}}}
                child {node [draw=none,minimum width=1em] {\large \textbf{\vdots}}}
            }
            child {node (L3b) {\textcolor{DeepRed}{AG}\textcolor{OliveGreen}{A}\textcolor{TealBlue}{C}TACT\\\textcolor{OliveGreen}{A}TC\textcolor{DarkOrchid}{CT}G\textcolor{DarkOrchid}{CT}\\\textcolor{OliveGreen}{A}A\textcolor{DarkOrchid}{C}TG\textcolor{DarkOrchid}{TCT}\\ fs(s)=\textcolor{OliveGreen}{1}+min(\textcolor{DarkOrchid}{4},\textcolor{DarkOrchid}{4})}
            [level distance=5.5em]
                child {node [draw=none, minimum width=1em] {\huge \textcolor{DeepRed}{\XSolidBrush} }}
            }
        }
        child {node (L2b) {\textcolor{DeepRed}{A}\textcolor{OliveGreen}{G}\textcolor{TealBlue}{A}CTACT\\ATCCT\textcolor{OliveGreen}{G}\textcolor{DarkOrchid}{CT}\\ACCT\textcolor{OliveGreen}{G}\textcolor{DarkOrchid}{TCT}\\ fs(s)=\textcolor{OliveGreen}{1}+min(\textcolor{DarkOrchid}{3},\textcolor{DarkOrchid}{4})}
            [level distance=5.5em]
            child {node [draw=none, minimum width=1em] {\huge \textcolor{DeepRed}{\XSolidBrush} }}
        }
    }
    child {node (R1) {\textcolor{OliveGreen}{A}\textcolor{TealBlue}{G}ACTACT\\\textcolor{OliveGreen}{A}TC\textcolor{DarkOrchid}{CT}G\textcolor{DarkOrchid}{CT}\\\textcolor{OliveGreen}{A}\textcolor{DarkOrchid}{AC}TG\textcolor{DarkOrchid}{TCT}\\ fs(s)=\textcolor{OliveGreen}{1}+min(\textcolor{DarkOrchid}{4},\textcolor{DarkOrchid}{5})}
        [sibling distance=9em]
        [level distance=6em]
        child {node (R2a) {\textcolor{OliveGreen}{A}\textcolor{DeepRed}{G}\textcolor{TealBlue}{A}CTACT\\\textcolor{OliveGreen}{A}TC\textcolor{DarkOrchid}{CT}G\textcolor{DarkOrchid}{CT}\\\textcolor{OliveGreen}{A}\textcolor{DarkOrchid}{AC}TG\textcolor{DarkOrchid}{TCT}\\ fs(s)=\textcolor{OliveGreen}{1}+min(\textcolor{DarkOrchid}{4},\textcolor{DarkOrchid}{5})}
            child {node (R3a) {\textcolor{OliveGreen}{A}\textcolor{DeepRed}{GA}\textcolor{TealBlue}{C}TACT\\\textcolor{OliveGreen}{A}TC\textcolor{DarkOrchid}{CT}G\textcolor{DarkOrchid}{CT}\\\textcolor{OliveGreen}{A}A\textcolor{DarkOrchid}{C}TG\textcolor{DarkOrchid}{TCT}\\ fs(s)=\textcolor{OliveGreen}{1}+min(\textcolor{DarkOrchid}{4},\textcolor{DarkOrchid}{4})}
                [level distance=4.2em]
                [sibling distance=9em]
                child {node [draw=none,minimum width=1em] {\large \textbf{\vdots}}}
                child {node [draw=none,minimum width=1em] {\large \textbf{\vdots}}}
            }
        }
        child {node (R2b) {\textcolor{OliveGreen}{AG}\textcolor{TealBlue}{A}CTACT\\\textcolor{OliveGreen}{A}TCCT\textcolor{OliveGreen}{G}\textcolor{DarkOrchid}{CT}\\\textcolor{OliveGreen}{A}ACT\textcolor{OliveGreen}{G}\textcolor{DarkOrchid}{TCT}\\ fs(s)=\textcolor{OliveGreen}{2}+min(\textcolor{DarkOrchid}{2},\textcolor{DarkOrchid}{3})}
            child {node (R3b) {\textcolor{OliveGreen}{AG}\textcolor{DeepRed}{A}\textcolor{TealBlue}{C}TACT\\\textcolor{OliveGreen}{A}TCCT\textcolor{OliveGreen}{G}\textcolor{DarkOrchid}{CT}\\\textcolor{OliveGreen}{A}ACT\textcolor{OliveGreen}{G}\textcolor{DarkOrchid}{TCT}\\ fs(s)=\textcolor{OliveGreen}{2}+min(\textcolor{DarkOrchid}{2},\textcolor{DarkOrchid}{3})}
                [level distance=4.2em]
                [sibling distance=8em]
                child {node [draw=none,minimum width=1em] {\large \textbf{\vdots}}}
                child {node [draw=none,minimum width=1em] {\large \textbf{\vdots}}}
            }
        }
    };

\draw[->, decorate, very thick, color=gray!70] (L2b.east) .. controls +(south:5em) and +(south:3.6em) .. (R2b.west);
\draw[->, decorate, very thick, color=gray!70] (L3b.east) .. controls +(south:2em) and +(south:2em) .. (R3a.west);

\end{tikzpicture}

\vspace{-0.15in}
\caption{
An example showing the progression of beam-guided ordering generator. The snake edges between the nodes denote that some nodes can be pruned when another node can always provide better results than the current node. The score of a node is the sum of the characters matched with all sequences and minimum of the characters matched in the pairwise MLCS between unmatched suffix of primary and other sequences.
}
\label{fig:beam}

\vspace{-0.15in}
\end{figure*}

\subsection{Beam-Guided Scoring and Upper Bound Formulation}
Beam search acts as a heuristic estimator of global importance. Instead of exhaustively exploring all possible subsequences, it maintains a limited number of promising candidates as it scans the primary sequence from left to right. At each position $i$, the beam branches into two possibilities: skipping the current character or keeping the current character if the resulting partial subsequence remains common across all other input sequences. Each candidate state is evaluated using a scoring function~\cite{djukanovic2021exact}:
$
f(s) = \ell + UB(i, c),
$
where $\ell$ is the number of characters already added (the current subsequence length), $i$ marks the current position in the primary sequence, $c$ represents cursor positions in the other sequences, and $UB(i, c)$ denotes an upper bound estimating how many additional matches can still be found from that point onward. This upper bound can be computed tightly using suffix-LCS tables. After evaluating both branches, the beam keeps only the top-$W$ highest-scoring states, ensuring that the search focuses on subsequences that are both valid and structurally promising.

Let the beam be at index $i$ in the primary sequence $P$, and let $c = (c_0, c_1, \ldots, c_{N-1})$ denote the cursor positions of the last matched characters in the other sequences $S_0, S_1, \ldots, S_{N-1}$. For each non-primary sequence $S_k$, a suffix-LCS table $L^{(k)}[p,q]$ is precomputed, where each entry represents the length of the longest common subsequence between the suffixes $P[p:]$ and $S_k[q:]$. During beam traversal, the remaining potential extension length is given by:
$
UB(i,c) = \min {}_k L^{(k)}[\,i,\, c_k{+}1\,].
$
This provides a reliable upper estimate of how many characters can still be matched simultaneously across all sequences. Since these tables are computed once, evaluating $UB(i,c)$ is efficient while still providing strong global guidance. The scoring function $f(s) = \ell + UB(i,c)$ thus balances the progress already made with the remaining potential, keeping the beam focused toward global optimal regions rather than local optima. An example of beam traversal is shown in Fig.~\ref{fig:beam}.

Once the beam traversal is complete, it produces several high-quality candidate paths. From these, the algorithm computes a \textit{survival frequency} for each index how often that character was included among the top-$T$ beam paths. This frequency serves as a statistical measure of the index's global importance. Positions with high survival counts are characters that persistently appear in strong subsequences and therefore should be prioritized. The indices are sorted in descending order of survival as high-survival characters are added first. 

\subsection{Primary Sequence Selection}
We introduce a primary sequence selection mechanism that identifies which sequences are most
likely to provide the best solution quality . For each sequence set under evaluation, we compute a
similarity-based ranking restricted to that set, that measures how representative each sequence is.
To construct this ranking efficiently, we extract a fixed-length prefix from each sequence and
estimate pairwise similarity through random sampling. Instead of evaluating all $N(N-1)/2$
possible pairs, a subset of sequence pairs is selected, and for each pair we compute the LCS
between their prefixes and normalize this value to obtain a comparable similarity score. By
averaging the similarity contributions we obtain a representativeness score for every sequence,
and these scores form the basis for ranking the sequences. Sequences with higher scores are
treated as stronger primary candidates and appear at the top of the ranking. Empirically, the
ranking is highly predictive of which sequence should be chosen as the primary. Across all three datasets used, the sequence that produces the longest MLCS appears within the “Top-3” ranked candidates in about $80$\% of cases, and in all remaining cases within the “Top-10”. As the
strongest primaries reliably fall within the highest-ranked candidates, ARP can limit evaluation to this small group and still achieve the maximal MLCS.

\begin{algorithm}[!t]
\caption{Primary Sequence Selection.}
\label{alg:primary_select}
\begin{algorithmic}[1]
\Require $S$: set of input sequences, $\ell$: fixed prefix length
\Ensure $P$: selected primary sequence

\Function{\sc PrimarySelect}{$S$}
    \State $N \leftarrow |S|$
    \State For each $S_i \in S$, extract prefix $P_i \leftarrow \text{prefix}(S_i,\ell)$
        
        \Comment{Fixed-length prefix for efficient similarity estimation}
    \State Initialize $\textit{score}[i] \leftarrow 0$ for all $i \in \{1,\dots,N\}$
    \State Randomly select a subset $\mathcal{R}$ of distinct pairs $(i,j)$ from the $\frac{N(N-1)}{2}$ possible pairs
    \For{each $(i,j) \in \mathcal{R}$}
        \State $L \leftarrow \text{LCSLen}(P_i, P_j)$
        \State $sim \leftarrow \dfrac{L}{\min(|P_i|,|P_j|)}$
            \Comment{Normalized prefix similarity}
        \State $\textit{score}[i] \leftarrow \textit{score}[i] + sim$
        \State $\textit{score}[j] \leftarrow \textit{score}[j] + sim$
    \EndFor
    \For{$i \leftarrow 1$ \textbf{to} $N$}
        \State $\textit{avg}[i] \leftarrow \dfrac{\textit{score}[i]}{(N-1)}$
            \Comment{Representative score}
    \EndFor
    \State Rank sequences by $\textit{avg}[i]$ in descending order
    \State $P \leftarrow$ top k highest-ranked sequences 
    \Comment{Selected primary sequence candidates}
    \State \Return $P$
\EndFunction
\end{algorithmic}
\end{algorithm}

\section{Experimental Results}
In this section we experimentally evaluate the effectiveness of the ARP algorithm. We first describe the datasets used and then experimentally compare our approach with related works.

\subsection{Datasets Used}  
The experiments are performed using five benchmark datasets: ACO-Random, ACO-Rat, and ACO-Virus~\cite{shyu2009finding} , BB~\cite{blum2007probabilistic} and ES~\cite{easton2008large}. ACO datasets include both DNA sequences ($|\Sigma|=4$) and Protein sequences ($|\Sigma|=20$). Each sequence has a fixed length of 600, and the number of sequences per instance ranges from 10 to 200. 
The BB dataset\cite{blum2007probabilistic}, originally introduced by Blum and Blesa, consists of structured synthetic instances generated using a base-string deletion model. For each configuration defined by alphabet size $|\Sigma|$, number of sequences $N$, and string length $\ell$, a base string of length $\ell$ is first randomly generated over $\Sigma$. Each of the $N$ sequences is then derived by independently deleting characters from this base string with a fixed probability (0.1), ensuring an underlying shared structure among sequences. In our experiments, we consider alphabet sizes $|\Sigma| \in \{2,4,8,24\}$, string length $\ell = 1000$, and $N \in \{10,100\}$, as shown in Table~\ref{tab:bbes}. For each parameter configuration, 10 independent instances are evaluated, and the reported results correspond to the average MLCS length over these 10 instances.
The ES dataset~\cite{easton2008large}, proposed by Easton and Singireddy, consists of fully random synthetic instances. For each combination of alphabet size $|\Sigma|$, number of sequences $N$, and string length $\ell$, sequences are generated independently by sampling each character uniformly from $\Sigma$. 
In our evaluation, we consider alphabet sizes ranging from 2 to 100, string lengths from 1000 to 5000, and $N \in \{10,50,100\}$ sequences per instance. Each configuration contains 50 independent instances, and all reported values are averaged over these 50 instances.
\subsection{Algorithms Compared} 
In our evaluation we compare the following algorithms:
\begin{itemize}
    \item ARP is our main algorithm whose default configuration is tuned for quality and has the default beam width of 2000 as quality improved as beam width was increased till 2000. To show benefits of replacement and prioritization, we also present results for AOnly that uses addition only and is our fastest configuration. The MAX depth used by our Replacement algorithm of ARP was also varied and our evaluation shows that the MAX depth of 2 gave the best results (see Figure 3). Therefore, results presented are for the MAX depth of 2. 
    \item BNMAS~\cite{huang2004fast} is a fast heuristic that was implemented following the details provided in the paper.
    \item UB-HH~\cite{abdi2023longest} is a powerful meta-heuristic that was evaluated using a beam width of 600, which corresponds to their high-quality scenario and represents the best configuration reported in their original paper. 
\end{itemize}
In each test, the reported ARP results are based on the best choice for the primary sequence.  All implementations are developed in Python to ensure consistency.

\newpage

\begin{table}[!t]

\centering
\normalsize
\caption{Comparison of solution quality and runtime (in seconds) for DNA ($|\Sigma|=4$) and protein ($|\Sigma|=20$) sequence instances. ARP/UB-HH give solution quality and T\_ARP/T\_UB-HH the time taken to generate the best solution.}
\label{tab:aco}
\vspace{0.05in}

\setlength{\tabcolsep}{3pt}
\renewcommand{\arraystretch}{1.0}   
\begin{tabular*}{\linewidth}{@{\extracolsep{\fill}}>{\centering\arraybackslash}p{0.5cm}rrrrrrrrr@{}}
\toprule
 & $|\Sigma|$ & $\ell$ & $N$ & BNMAS & AOnly & 
 ARP & UB-HH & T\_ARP & T\_UB-HH \\
\midrule
\multirow{16}{*}{\rotatebox{90}{\normalsize\bfseries ACO-Random}}
 & 4  & 600 & 10  & 144 & 150 & 
 219 & 218 & 5.64 & 11.7 \\
 & 4  & 600 & 20  & 140 & 143 &
 191 & 193& 6.7 &12.83 \\
 & 4  & 600 & 40  & 129 & 139 &
 172 & 175 &15.39 &9.29\\
 & 4  & 600 & 60  & 132 & 137 & 
 165 & 168&11.4 &18.58\\
 & 4  & 600 & 80  & 131 & 136 & 
 160 & 162 &13.95 & 21.95\\
 & 4  & 600 & 100 & 131 & 136 & 
 156 & 159 &16.4 &24.63\\
 & 4  & 600 & 150 & 122 & 136 & 
 150 & 150 & 23.49 & 33.27\\
 & 4  & 600 & 200 & 124 & 133 &
 150 & 151&29.98 &40.91 \\
\addlinespace[2pt]
 & 20 & 600 & 10  & 30 & 33 & 
 57 & 62 & 5.33& 14.9 \\
 & 20 & 600 & 20  & 26 & 32 & 
 45 & 47 & 6.75 & 13.48\\
 & 20 & 600 & 40  & 26 & 26 & 
 36 & 39 & 8.39 &14.31 \\
 & 20 & 600 & 60  & 26 & 27 & 
 33 & 35& 10.4 & 16.53 \\
 & 20 & 600 & 80  & 26 & 25 & 
 31 & 33 &12.5 & 18.85\\
 & 20 & 600 & 100 & 22 & 25 & 
 31 & 32 & 14.51 & 21.73 \\
 & 20 & 600 & 150 & 23 & 24 & 
 28 & 29 & 19.53 & 26.93\\
 & 20 & 600 & 200 & 21 & 23 & 
 28 & 28 & 23.9 & 32.04 \\
\midrule
\multirow{16}{*}{\rotatebox{90}{\normalsize\bfseries ACO-Rat}}

 & 4 & 600 & 10 & 91 & 137 &
 206 & 204 &5.41 &11.19\\
 & 4 & 600 & 20 & 126 & 136 &
 169 & 171 & 7.93& 11.1\\
 & 4 & 600 & 40 & 118 & 130 & 
 152 & 153 &9.88 & 12.23\\
 & 4 & 600 & 60 & 104 & 120 &
 150 & 152 & 11.79 & 16.47\\
 & 4 & 600 & 80 & 73 & 109 & 
 137 & 140 &10.51 & 60.16 \\
 & 4 & 600 & 100 & 93 & 111 &
 135 & 137 & 31.89 & 77.54 \\
 & 4 & 600 & 150 & 62 & 108 & 
 127 & 129 & 56.28& 105.61 \\
 & 4 & 600 & 200 & 64 & 110 & 
 120 & 124 & 98.77 & 144.5\\
\addlinespace[2pt]
 & 20 & 600 & 10 & 40 & 42 &
 68 & 70 & 37.9 & 61.51 \\
 & 20 & 600 & 20 & 33 & 34 & 
 53 & 54 &30.88 & 68.94 \\
 & 20 & 600 & 40 & 30 & 31 & 
 48 & 50 &50.09& 77.19\\
 & 20 & 600 & 60 & 30 & 31 & 
 46 & 49 &78.872&90.86 \\
 & 20 & 600 & 80 & 32 & 32 &
 41 & 44 &117.5& 129.08 \\
 & 20 & 600 & 100 & 30 & 35 & 
 40 & 40 & 150.04&182\\
 & 20 & 600 & 150 & 25 & 26 &
 36 & 38& 96.1& 216.04 \\
 & 20 & 600 & 200 & 25 & 26 &
 34 & 35 & 107.1&226\\
\midrule
\multirow{16}{*}{\rotatebox{90}{\normalsize\bfseries ACO-Virus}}
 & 4 & 600 & 10 & 126 & 142 & 
 226 & 223 &5.47& 12.17\\
 & 4 & 600 & 20 & 113 & 136 &
 189 & 190&7.15&12.46 \\
 & 4 & 600 & 40 & 112 & 129 &
 168 & 171& 9.87 & 15.08 \\
 & 4 & 600 & 60 & 112 & 127 & 
 164 & 166 & 11.3&18.07\\
 & 4 & 600 & 80 & 104 & 126 & 
 159 & 160 &14.75 & 21.56\\
 & 4 & 600 & 100 & 104 & 126 & 
 157 & 159& 17.74 & 25.03 \\
 & 4 & 600 & 150 & 100 & 128 & 
 159 & 157 & 24.58&33.92\\
 & 4 & 600 & 200 & 100 & 128 &
 153 & 155& 32.84 &42.37 \\
\addlinespace[2pt]
 & 20 & 600 & 10 & 37 & 39 & 
 78 & 75 & 5.23& 15.7\\
 & 20 & 600 & 20 & 32 & 33 & 
 62 & 60&63.3&87.47 \\
 & 20 & 600 & 40 & 31 & 33 & 
 50 & 51&76.09&90.89 \\
 & 20 & 600 & 60 & 31 & 32 & 
 45 & 44 &87.79&126.41\\
 & 20 & 600 & 80 & 29 & 30 & 
 46 & 46&80.18&176.37 \\
 & 20 & 600 & 100 & 27 & 29 &
 42 & 44&122.96&225.86 \\
 & 20 & 600 & 150 & 27 & 30 & 
 45 & 45 &155.17&267.04\\
 & 20 & 600 & 200 & 29 & 30 & 
 44 & 44&219.64 &381.08 \\
\bottomrule
\end{tabular*}
\vspace{-0.15in}
\end{table}

\begin{table}[!t]
\centering
\normalsize
\caption{Comparison of solution quality and runtime (in seconds) on BB and ES datasets with varying alphabet size ($|\Sigma|=$ 2 to 100), long string lengths ($\ell=$ 1000 to 5000), and 10 to 100 strings.}
\label{tab:bbes}
\vspace{0.1in}
\setlength{\tabcolsep}{3pt}
\renewcommand{\arraystretch}{1.0}
\setlength{\tabcolsep}{3pt}
\renewcommand{\arraystretch}{1.0}   
\begin{tabular*}{\linewidth}{@{\extracolsep{\fill}}>{\centering\arraybackslash}p{0.5cm}rrrrrrrrr@{}}
\toprule
 & $|\Sigma|$ & $\ell$ & $N$ & BNMAS & AOnly & 
 ARP & UB-HH& T\_ARP & T\_UB-HH \\ 
\midrule
\multirow{12}{*}{\rotatebox[origin=c]{90}{\normalsize\bfseries ES}}
 & 2  & 1000 & 10  & 483.86 & 492.22 &
 609.29 & 611.48 & 20.04&17.37\\
 & 2  & 1000 & 50  & 467.42 & 475.86 & 
 535.30 & 538.22&27.67 &31.08\\
 & 2  & 1000 & 100 & 463.34 & 462.02 &
 516.75 & 519.84&47.22&96.86 \\
\addlinespace[2pt]
 & 10 & 1000 & 10  & 106.24 & 112.28 & 
 194.32 & 200.60 &11.1&27.16\\
 & 10 & 1000 & 50  & 95.30  & 96.80  &
 131.54 & 136.22&18.4& 33.89\\
 & 10 & 1000 & 100 & 91.44  & 94.48  &
 119.20 & 123.28&29.04& 51.27\\
\addlinespace[2pt]
 & 25 & 2500 & 10  & 101.50 & 105.04 &
 228.70 & 231.92 &46&83.85\\
 & 25 & 2500 & 50  & 89.50  & 91.20  &
 132.56 & 139.38 &70& 98.01\\
 & 25 & 2500 & 100 & 84.76  & 87.44  &
 120.40 & 122.78 &94.6&133.74\\
\addlinespace[2pt]
 & 100 & 5000 & 10 & 52.70  & 54.68  &
 136.50 & 140.78 &184&190.59\\
 & 100 & 5000 & 50 & 41.70  & 42.88  &
 68.00  & 71.28 &197& 204.23\\
 & 100 & 5000 & 100& 39.72  & 41.50  &
 56.50  & 60.12 &278&284.96\\
\midrule
\multirow{8}{*}{\rotatebox[origin=c]{90}{\bfseries BB}}
 & 2  & 1000 & 10  & 410.10 & 448.00 & 
 677.10 & 634.90 &18.2&18.7 \\
 & 2  & 1000 & 100 & 320.00 & 425.80 & 
 552.55 & 562.00&45.78 &48.42 \\
\addlinespace[2pt]
 & 4  & 1000 & 10  & 221.40 & 227.80 &
 449.30 & 453.00 &13.09& 26.36\\
 & 4  & 1000 & 100 & 115.50 & 218.80 &
 360.00 & 372.30& 50.48 & 63.48 \\
\addlinespace[2pt]
 & 8  & 1000 & 10  & 130.10 & 134.60 &
 337.60 & 341.40 & 10.69&43.62\\
 & 8  & 1000 & 100 & 75.70  & 111.80 &
 238.30 & 247.90&66.53&84.5 \\
\addlinespace[2pt]
 & 24 & 1000 & 10  & 53.20  & 88.20  & 
 384.40 & 281.80&18.35&104.67 \\
 & 24 & 1000 & 100 & 33.40  & 43.50  &
 149.60 & 134.10 &98.41&133.06\\
\bottomrule
\end{tabular*}
\end{table}

\vspace{1in}
\begin{table}[!t]
\centering
\normalsize
\caption{Comparison of AOnly against BNMAS on the ACO-Random, ACO-Rat, ACO-Virus, BB and ES datasets, with solution quality and runtime (in seconds) totalled over all configurations of each dataset. Speedup gives the ratio of T\_BNMAS to T\_AOnly and Gain the relative increase in solution quality of AOnly over BNMAS.}
\label{tab:aonly-vs-bnmas}
\vspace{0.1in}
\setlength{\tabcolsep}{3pt}
\renewcommand{\arraystretch}{1.0}
\begin{tabular*}{\linewidth}{@{\extracolsep{\fill}}lrrrrrr@{}}
\toprule
Dataset & T\_BNMAS & T\_AOnly & Speedup & BNMAS & AOnly & Gain \\
\midrule
ACO-Random & 1.85  & 12.29  & 6.7$\times$  & 1253.00 & 1325.00 & $+6\%$  \\
ACO-Rat    & 1.59  & 19.83  & 12.4$\times$ & 976.00  & 1218.00 & $+25\%$ \\
ACO-Virus  & 0.21  & 0.46   & 2.2$\times$  & 1114.00 & 1298.00 & $+17\%$ \\
\addlinespace[2pt]
BB         & 0.70  & 12.48  & 17.8$\times$ & 1359.40 & 1698.50 & $+25\%$ \\
ES         & 14.99 & 903.82 & 60.3$\times$ & 2117.48 & 2156.40 & $+2\%$  \\
\bottomrule
\end{tabular*}
\end{table}
\begin{figure}[t]
    \centering
   
    \includegraphics[width=0.55\linewidth]{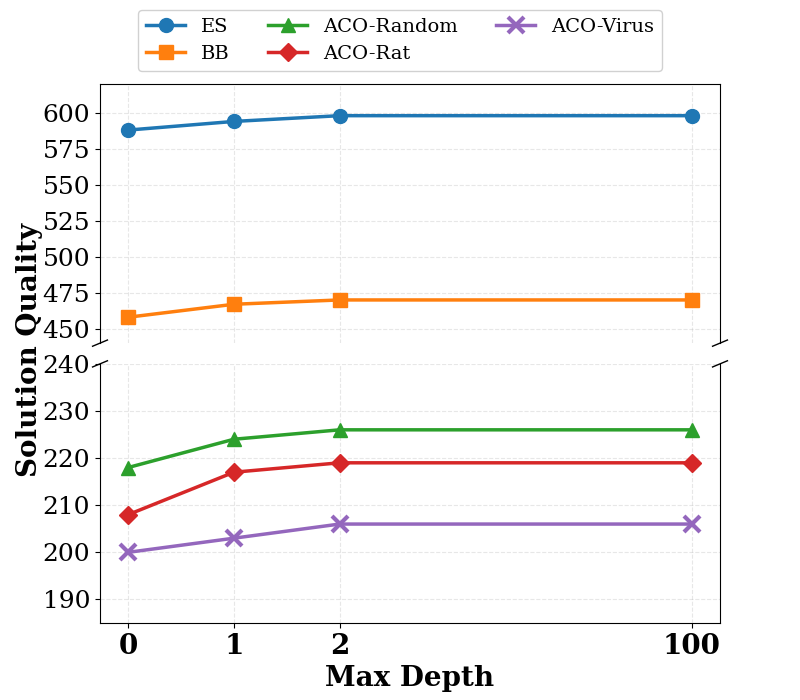}

    \vspace{-0.15in}
    \caption{Sensitivity analysis of maximum depth 0, 1, 2 and 100 is performed for the ARP algorithm across ACO-Random, ACO-Rat, and ACO-Virus, BB and ES datasets. The solution quality increases from Max Depth = 0 to 2 and remains nearly constant thereafter.}
    \label{fig:sensitivity}
    
   \vspace{-0.15in}
\end{figure}
\newpage
\null
\newpage
\subsection{Results}

The results for common subsequence lengths and time found using various algorithms for the ACO-Random, ACO-Rat, and ACO-Virus, BB and ES datasets are presented in Tables~\ref{tab:aco} and \ref{tab:bbes}. 
Table~\ref{tab:bbes} presents the average solution quality for each parameter configuration, computed over 50 independent instances for ES and 10 independent instances for BB. Unlike Table 2 where $\ell$ = 600, these experiments involve substantially longer inputs, with $\ell$ ranging from 1000 to 5000, resulting in a significantly larger search space and increased computational complexity. For each parameter configuration defined by $|\Sigma|$, $\ell$ and $N$, we ran our algorithm independently on every instance and report the average solution quality across those runs, which explains the presence of decimal values in the table. Due to the larger input sizes and varying alphabet configurations, these experiments evaluate ARP under more demanding and diverse settings. 
By examining the above results in Tables~\ref{tab:aco}, \ref{tab:bbes}, and \ref{tab:aonly-vs-bnmas} we make the following observations:

\begin{itemize}
 \item
 \textsf{AOnly vs. BNMAS.}
The AOnly version, which uses only addition
but no replacement or prioritization, is a lightweight approach. In Table~\ref{tab:aonly-vs-bnmas}, we can see that
although BNMAS is faster than AOnly, AOnly substantially
outperforms BNMAS in terms of the lengths of the common
subsequences found.

 \item 
 \textsf{ARP vs. UB-HH.}
 On the ACO (Table~\ref{tab:aco}) and ES (Table~\ref{tab:bbes}) datasets, which consists of fully random instances, ARP closely approaches UB-HH in terms of solution quality across all settings. Although the input length in Table~\ref{tab:bbes} are much larger, the differences between ARP and UB-HH remain relatively small across most configurations, indicating that ARP remains competitive even for larger-scale instances. On the structured BB dataset, ARP achieves particularly strong improvements and, in several configurations, matches or exceeds UB-HH.
 ARP achieves solution quality comparable to UB-HH while outperforming it in runtime T\_ARP by 1.1$\times$--1.7$\times$ faster relative to T\_UB-HH across all five datasets. As mentioned earlier, the results discussed above were obtained by setting the MAX depth to 2. Finally, the sensitivity study results justifying this choice of MAX depth of 2 are shown in Figure~\ref{fig:sensitivity}. The analysis shows that depth 2 captures most of the improvement in solution quality, making it a practical setting for all reported results.
\end{itemize}

  
\section{Conclusions}

In this work a new formulation for the Multiple Longest Common Subsequence problem, ARP, was introduced to systematically explore subset and complement partitions of a primary sequence. By ordering characters for inclusion in a solution, and replacing subsolutions with alternatives, the ability to identify longer common subsequences is greatly enhanced. 
Our fastest Addition version , which is A(0) that does not order characters, outperforms the BNMAS heuristic in subsequence length across five datasets. 
ARP that orders characters, matches hyperheuristic UB-HH performance in the common subsequence lengths found across all five datasets.
These results show that ARP is a practical algorithm that matches or improves upon the performance of other techniques.

\bibliographystyle{unsrt} 
\bibliography{reference}
\section{Funding}
This work is supported in part by National Science Foundation Grants CCF-2512416 and CCF-2226448 to the UCR.
\section{Conflict of interest}
None declared
\section{Data availability}
The benchmark datasets and source code underlying this article are
available at \url{https://github.com/ftumpa001/ARP-MLCS} and archived
at \url{https://doi.org/10.5281/zenodo.19168760}

\end{document}